\documentclass[aps,superscriptaddress,preprintnumbers,showpacs]{revtex4}
\usepackage{amssymb}
\usepackage{graphicx}
\usepackage{dcolumn}
\usepackage{bm}
\usepackage{amsmath}

\begin{document}

\newcommand*{\PKU}{School of Physics and State Key Laboratory of Nuclear Physics and
Technology, Peking University, Beijing 100871, China}\affiliation{\PKU}
\newcommand*{\chep}{Center for High Energy Physics, Peking University, Beijing 100871, China}\affiliation{\chep}

\title{Parametrization of fermion mixing matrices in Kobayashi-Maskawa form}

\author{Nan Qin}\affiliation{\PKU}
\author{Bo-Qiang Ma}\email{mabq@pku.edu.cn}\affiliation{\PKU}\affiliation{\chep}

\begin{abstract}
Recent works show that the original Kobayashi-Maskawa (KM) form of
fermion mixing matrix exhibits some advantages, especially when discussing
problems such as unitarity boomerangs and maximal CP violation hypothesis.
Therefore, the KM form of fermion mixing matrix is systematically studied in this paper.
Starting with a general triminimal expansion of the KM matrix, we discuss the triminimal and
Wolfenstein-like parametrizations with different basis matrices in detail.
The quark-lepton complementarity relations play an important role in our discussions
on describing quark mixing and lepton mixing in a unified way.

\end{abstract}


\maketitle

\section{Introduction}

As it is well known, the mixing between different generations of
fermions is one of the most interesting issues in particle physics. For
quarks, the mixing matrix is described by the
Cabibbo\cite{cabibbo}-Kobayashi-Maskawa\cite{km}(CKM) matrix $V_{\rm
CKM}$, and in the lepton sector, it is described by the
Pontecorvo\cite{pontecorvo}-Maki-Nakawaga-Sakata\cite{mns} (PMNS)
matrix $U_{\rm PMNS}$,
\begin{eqnarray}
V_{\rm CKM}=\left(
  \begin{array}{ccc}
    V_{ud} & V_{us} & V_{ub} \\
    V_{cd} & V_{cs} & V_{cb} \\
    V_{td} & V_{ts} & V_{tb} \\
  \end{array}
\right), \quad U_{\rm PMNS}=\left(
  \begin{array}{ccc}
    U_{e1}    & U_{e2}    & U_{e3}    \\
    U_{\mu1}  & U_{\mu2}  & U_{\mu3}  \\
    U_{\tau1} & U_{\tau2} & U_{\tau3} \\
  \end{array}
\right).\nonumber
\end{eqnarray}
Before more underlying theory of the origin of the mixing is found,
parametrizing the mixing matrices properly is helpful to
understanding the mixing pattern and search for deviations from the
standard model both theoretically and experimentally. A commonly
used form of the fermion mixing matrix is the standard parametrization
proposed by Chau and Keung (CK)~\cite{ck}
\begin{eqnarray}
V_{\rm CK}&=&\left(
\begin{array}{ccc}
1 & 0 & 0           \\
0 & c_{23} & s_{23} \\
0 & -s_{23} & c_{23}\\
\end{array}
\right)\left(
\begin{array}{ccc}
c_{13} & 0 & s_{13}e^{-i\delta_{\rm CK}}\\
0 & 1 & 0 \\
-s_{13}e^{i\delta_{\rm CK}} & 0 & c_{13}\\
\end{array}
\right)\left(
\begin{array}{ccc}
c_{12}& s_{12} & 0\\
-s_{12} & c_{12} & 0 \\
0 & 0 & 1\\
\end{array}
\right)\nonumber\\
&=&\left(
\begin{array}{ccc}
c_{12}c_{13} & s_{12}c_{13} & s_{13}e^{-i\delta_{\rm CK}}           \\
-s_{12}c_{23}-c_{12}s_{23}s_{13}e^{i\delta_{\rm CK}} &
c_{12}c_{23}-s_{12}s_{23}s_{13}e^{i\delta_{\rm CK}}  & s_{23}c_{13} \\
s_{12}s_{23}-c_{12}c_{23}s_{13}e^{i\delta_{\rm CK}}  &
-c_{12}s_{23}-s_{12}c_{23}s_{13}e^{i\delta_{\rm CK}} & c_{23}c_{13}
\end{array}
\right),\label{CK}
\end{eqnarray}
where $s_{ij}=\sin\theta_{ij}$ and $c_{ij}=\cos\theta_{ij}$
$(i,j=1,2,3)$ are the rotation angles, and $\delta_{\rm CK}$ is the
CP-violating phase in the CK parametrization.

Recently, it has been pointed out in many
works~\cite{boomerang,recentwork,koide} that the original
Kobayashi-Maskawa (KM)~\cite{km} matrix is convenient when
discussing problems such as unitarity boomerangs~\cite{boomerang}
and maximal CP violation hypothesis~\cite{maximalcp}. The original
KM mixing matrix is given by
\begin{eqnarray}
V_{\rm KM}&=&\left(
  \begin{array}{ccc}
    1 & 0   & 0    \\
    0 & c_2 & -s_2 \\
    0 & s_2 & c_2  \\
  \end{array}
\right)\left(
  \begin{array}{ccc}
    c_1 & -s_1 & 0 \\
    s_1 & c_1  & 0 \\
    0   & 0    & e^{i\delta_{\rm KM}} \\
  \end{array}
\right)\left(
  \begin{array}{ccc}
    1 & 0   & 0    \\
    0 & c_3 & s_3  \\
    0 & s_3 & -c_3 \\
  \end{array}
\right)\nonumber\\
&=&\left(
\begin{array}{ccc}
c_1     & -s_1c_3                     & -s_1s_3           \\
s_1c_2  & c_1c_2c_3-s_2s_3e^{i\delta_{\rm KM}} & c_1c_2s_3+s_2c_3e^{i\delta_{\rm KM}} \\
s_1s_2  & c_1s_2c_3+c_2s_3e^{i\delta_{\rm KM}} &
c_1s_2s_3-c_2c_3e^{i\delta_{\rm KM}}
\end{array}
\right)\;,\label{KM}
\end{eqnarray}
in which $s_i=\sin\theta_i$, $c_i=\cos\theta_i$ $(i=1,2,3)$ are Euler
angles, and $\delta_{\rm KM}$ is the CP-violating phase in the KM
parametrization.

If neutrinos are of the Majorana type, there should be an additional
diagonal matrix with two Majorana phases $P={\rm diag}(e^{i\alpha_1/2},e^{i\alpha_2/2},1)$
multiplied to Eqs.~(\ref{KM}) and (\ref{CK}) from the right. In this paper, we
consider the neutrinos as Dirac neutrinos, and the presentation of
formalisms for Majorana neutrinos can be derived straightforwardly
by including the additional phases. In the following, we omit
the subscript $\rm CK$ and $\rm KM$ since we deal with only the KM
form. We also denote parameters in the quark sector with superscript $Q$
and in the lepton sector with superscript $L$ if necessary (except for the
CP-violating phases in expressions with the consideration of concinnity,
and bearing in mind that $\delta^Q$ appears in $V_{\rm CKM}$ while
$\delta^L$ appears in $U_{\rm PMNS}$ ).

The magnitudes of the CKM matrix elements have been well determined
with~\cite{pdg}
\begin{eqnarray} \left(
  \begin{array}{ccc}
    0.97428\pm0.00015             & 0.2253\pm0.0007                & 0.00347^{+0.00016}_{-0.00012}               \\
    0.2252\pm0.0007               & 0.97345^{+0.00015}_{-0.00016}  & 0.0410^{+0.0011}_{-0.0007}      \\
    0.00862^{+0.00026}_{-0.00020} & 0.0403^{+0.0011}_{-0.0007}     & 0.999152^{+0.000030}_{-0.000045}
  \end{array} \right)\;.\label{ckmdata}
\end{eqnarray}
With $|V_{ud}|$, $|V_{ub}|$, $|V_{td}|$ and $|V_{cb}|$ as input parameters, one can easily get the ranges for the angle parameters as
\begin{eqnarray}
\theta_1^Q=0.2273_{-0.0003}^{+0.0011},\quad
\theta_2^Q=0.0383_{-0.0010}^{+0.0011},\quad
\theta_3^Q=0.0154_{-0.0006}^{+0.0008},\quad
\delta^Q={90.33^{\circ}}_{-4.57^{\circ}}^{+2.85^{\circ}}\;.\label{kmdata}
\end{eqnarray}
The last equation apparently implies that the KM phase convention is consistent with the maximal CP violation hypothesis.

For lepton mixing, the ranges for the PMNS matrix elements have been
also constrained by (at $3\sigma$ level)~\cite{gonzalez}
\begin{eqnarray} \left(
  \begin{array}{ccc}
    0.77 - 0.86      & 0.50 - 0.63      & 0.00 - 0.22      \\
    0.22 - 0.56      & 0.44 - 0.73      & 0.57 - 0.80      \\
    0.21 - 0.55      & 0.40 - 0.71      & 0.59 - 0.82
  \end{array} \right)\;.\label{pmnsdata}
\end{eqnarray}
Since the data are not accurate enough here, we do not calculate the parameters for leptons as what we do for quarks. Instead,
the numerical results are presented in Sec.~III and Sec.~IV where unified description of quark mixing
and lepton mixing are discussed.

When studying mixing, it is useful to parametrize the matrix
according to the hierarchical structure of the mixing to reveal more
physical information about the underlying theory. The Wolfenstein
parametrization for quarks is a famous example of this type, where
$V_{\rm CKM}$ is parametrized as~\cite{wolf}
\begin{eqnarray}
V_{\rm CKM}&=&\left( \begin{array}{ccc}
1-\frac{1}{2}\lambda^2&\lambda&A\lambda^3(\rho-i\eta)\\
-\lambda&1-\frac{1}{2}\lambda^2&A\lambda^2\\
A\lambda^3(1-\rho-i\eta)&-A\lambda^2&1
\end{array}\right)
+\mathcal {O}(\lambda^4)\;.\label{wolf}
\end{eqnarray}
The up-to-date fit for the Wolfenstein parameters gives~\cite{pdg}
\begin{eqnarray}
&&\lambda=0.2253\pm0.0007\;,
\quad\quad A=0.808_{-0.015}^{+0.022}\;,\nonumber\\
&&\rho(1-\lambda^2/2+\ldots)=0.132_{-0.014}^{+0.022}\;, \quad
\eta(1-\lambda^2/2+\ldots)=0.341\pm0.013\;.\label{wolfdata}
\end{eqnarray}

The Wolfenstein parametrization is actually an expansion of $V_{\rm
CKM}$ around the unit matrix basis with $\lambda$ as the expanding
parameter. In this type of parametrization, the choice of the
parameters and where to put them are arbitrary, making the meaning
of the parameters subtle to some extent. For example, the
CP-violating phase $\delta$ is not independent, i.e., it is
determined by two parameters $\eta$ and $\rho$ with
$\tan{\delta}=\eta/\rho$. Therefore it would be better to expand the
mixing matrix using small parameters with explicit physical meaning.
A good choice is the idea of triminimal
parametrization~\cite{triminimal} with an approximation as the basis
matrix to the lowest order. The triminimal expansion of the quark
and lepton mixing pointed out a new way to parametrize the mixing
matrix with all angle parameters small, and with the CP-violating
phase parameter free from others. The parameters are
completely determined when the basis matrix is chosen.

For quark mixing, the unit matrix is very simple while the matrix
suggested in Ref.~\cite{lisw} given by
\begin{eqnarray}
V'_0=\left(
        \begin{array}{ccc}
            \frac{\sqrt{2}+1}{\sqrt{6}} & \frac{\sqrt{2}-1}{\sqrt{6}} & 0 \\
           -\frac{\sqrt{2}-1}{\sqrt{6}} & \frac{\sqrt{2}+1}{\sqrt{6}} & 0 \\
            0                           & 0                           & 1
        \end{array}
        \right)\label{vnew}
\end{eqnarray}
is more close to experimental data so that they are both good
choices.

In lepton sector, it has been common to choose the bimaximal
matrix~\cite{bi} and/or the tri-bimaximal matrix~\cite{tri} as the
basis matrices
\begin{eqnarray}
U_{\rm{bi}}=\left(
              \begin{array}{ccc}
                1/\sqrt{2} & 1/\sqrt{2} & 0 \\
                -1/2       & 1/2        & 1/\sqrt{2} \\
                1/2        & -1/2       & 1/\sqrt{2} \\
              \end{array}
            \right),\quad
U_{\rm{tri}}=\left(
              \begin{array}{ccc}
              2/\sqrt{6}  & 1/\sqrt{3}  & 0          \\
              -1/\sqrt{6} & 1/\sqrt{3}  & 1/\sqrt{2} \\
              1/\sqrt{6}  & -1/\sqrt{3} & 1/\sqrt{2}
              \end{array}
            \right).
\end{eqnarray}
Although the former one is not favored by present experimental data
as the later one, it looks more symmetric with also a possible
connection with the unit basis in quark mixing~\cite{qlcparabi}. The
tri-bimaximal basis is very close to experimental data and can serve
as a good approximation for lepton mixing.

Although it seems that the mixing of quarks and leptons are
unrelated with each other, there indeed exist phenomenological
relations between mixing angles called quark-lepton complementarity
(QLC)~\cite{qlc}. For KM parameters, the QLC relations still
stand~\cite{recentwork,zheng}, 
\begin{eqnarray}
\theta_{1}^Q+\theta_{1}^L=\frac{\pi}{4},
\quad\theta_{2}^Q+\theta_{2}^L=\frac{\pi}{4},
\quad\theta_{3}^Q\sim\theta_{3}^L\sim 0\;.\label{kmqlc}
\end{eqnarray}
It has been discussed that the quark mixing matrix and the lepton
mixing matrix can be parametrized in a unified way with the QLC
relations~\cite{unified}. However, the discussions in Ref.~\cite{unified}
are based on the CK phase convention. Since the KM form of mixing matrices
is promoted in many works, a detailed study of it is necessary,
and unified parametrizations in KM phase convention may be helpful in both
theoretical and phenomenological studies.

The outline of this paper is as follows. In Sec.~II, the general
expressions of triminimal expansion of the KM matrix are presented.
In Sec.~III, we study the triminimal parametrization with unit
matrix and bimaximal matrix as the basis for quarks and leptons
respectively. Wolfenstein-like parametrizations are also discussed
and numerical results of the parameters are presented. In Sec.~IV,
triminimal and Wolfenstein-like parametrizations are discussed in
tri-bimaximal pattern. We show that the expansions converge much
faster in both quark sector and lepton sector. In both Sec.~III and
Sec.~IV, QLC relations play an important role in our discussions of
parametrizing $V_{\rm CKM}$ and $U_{\rm PMNS}$ in a unified way.
Finally, we present our conclusions in Sec.~V.

\section{The general results of triminimal expansion of KM matrix}

The idea of the triminimal parametrization~\cite{triminimal} is to
express a mixing angle in the mixing matrices as the sum of a zeroth
order angle $\theta^0$ and a small perturbation angle $\epsilon$ as
\begin{eqnarray}
\theta_1=\theta_1^0+\epsilon_1,\quad
\theta_2=\theta_2^0+\epsilon_2,\quad \theta_3=\theta_3^0+\epsilon_3.
\end{eqnarray}
With the deviations $\epsilon_i$, one can expand the mixing matrices
in powers of $\epsilon_i$ while different choices of $\theta_i^0$
lead to different matrices as the zeroth order of the expansion. Generally, to the
second order of $\epsilon_i$, the mixing matrix is expanded as
\begin{eqnarray}
V&=&\left( \begin{array}{ccc} c_1^0 & -s_1^0 c_3^0 & -s_1^0 s_3^0
\\s_1^0c_2^0 & c_1^0 c_2^0 c_3^0-s_2^0 s_3^0 e^{i\delta} & c_1^0 c_2^0 s_3^0+s_2^0 c_3^0 e^{i\delta}
\\s_1^0s_2^0&c_1^0s_2^0c_3^0+c_2^0s_3^0e^{i\delta}&c_1^0s_2^0s_3^0-c_2^0c_3^0e^{i\delta}
\end{array}\right)
+\epsilon_1\left( \begin{array}{ccc} -s_1^0 & -c_1^0c_3^0
&-c_1^0s_3^0
\\ c_1^0c_2^0 & -c_2^0c_3^0s_1^0 & -c_2^0s_1^0s_3^0 \\
c_1^0s_2^0 & -c_3^0s_1^0s_2^0 &-s_1^0s_2^0s_3^0
\end{array}\right)\nonumber\\
&+&\epsilon_2\left(\begin{array}{ccc}
0&0&0\\
-s_1^0s_2^0 &-c_1^0c_3^0s_2^0-c_2^0s_3^0e^{i\delta} &-c_1^0s_2^0s_3^0+c_2^0c_3^0e^{i\delta}\\
c_2^0s_1^0
&c_1^0c_2^0c_3^0-s_2^0s_3^0e^{i\delta}&c_1^0c_2^0s_3^0+c_3^0s_2^0e^{i\delta}
\end{array}\right)\nonumber\\
&+&\epsilon_3\left( \begin{array}{ccc}
0&s_1^0s_3^0&-c_3^0s_1^0\\
0&-c_1^0c_2^0s_3^0-c_3^0s_2^0e^{i\delta}&c_1^0c_2^0c_3^0-s_2^0s_3^0e^{i\delta}\\
0&-c_1^0s_2^0s_3^0+c_2^0c_3^0e^{i\delta}&c_1^0c_3^0s_2^0+c_2^0s_3^0e^{i\delta}
\end{array}\right)
+\frac{1}{2}\epsilon_1^2\left(\begin{array}{ccc}
-c_1^0&c_3^0s_1^0&s_1^0s_3^0\\
-c_2^0s_1^0&-c_1^0c_2^0c_3^0&-c_1^0c_2^0s_3^0\\
-s_1^0s_2^0&-c_1^0c_3^0s_2^0&-c_1^0s_2^0s_3^0
\end{array}\right)\nonumber\\
&+&\frac{1}{2}\epsilon_2^2\left(\begin{array}{ccc}
0&0&0\\
-c_2^0s_1^0&-c_1^0c_2^0c_3^0+s_2^0s_3^0e^{i\delta}&-c_1^0c_2^0s_3^0-c_3^0s_2^0e^{i\delta}\\
-s_1^0s_2^0&-c_1^0c_3^0s_2^0-c_2^0s_3^0e^{i\delta}&-c_1^0s_2^0s_3^0+c_2^0c_3^0e^{i\delta}
\end{array}\right)\nonumber\\
&+&\frac{1}{2}\epsilon_3^2\left(\begin{array}{ccc}
0&c_3^0s_1^0&s_1^0s_3^0\\
0&-c_1^0c_2^0c_3^0+s_2^0s_3^0e^{i\delta}&-c_1^0c_2^0s_3^0-c_3^0s_2^0e^{i\delta}\\
0&-c_1^0c_3^0s_2^0-c_2^0s_3^0e^{i\delta}&-c_1^0s_2^0s_3^0+c_2^0c_3^0e^{i\delta}
\end{array}\right)
+\epsilon_1\epsilon_2\left( \begin{array}{ccc}
0&0&0\\
-c_1^0s_2^0&c_3^0s_1^0s_2^0&s_1^0s_2^0s_3^0\\
c_1^0c_2^0&-c_2^0c_3^0s_1^0&-c_2^0s_1^0s_3^0
\end{array}\right)\nonumber\\
&+&\epsilon_2\epsilon_3\left( \begin{array}{ccc}
0&0&0\\
0&c_1^0s_2^0s_3^0-c_2^0c_3^0e^{i\delta}&-c_1^0c_3^0s_2^0-c_2^0s_3^0e^{i\delta}\\
0&-c_1^0c_2^0s_3^0-c_3^0s_2^0e^{i\delta}&c_1^0c_2^0c_3^0-s_2^0s_3^0e^{i\delta}
\end{array}\right)
+\epsilon_1\epsilon_3\left(\begin{array}{ccc}
0&c_1^0s_3^0&-c_1^0c_3^0\\
0&c_2^0s_1^0s_3^0&-c_2^0c_3^0s_1^0\\
0&s_1^0s_2^0s_3^0&-c_3^0s_1^0s_2^0
\end{array}\right)\nonumber\\
&+&\mathcal {O}(\epsilon_i^3)\;,\label{general}
\end{eqnarray}
where $s_i^0=\sin{\theta_i^0}$ and $c_i^0=\cos{\theta_i^0}$.

The rephasing invariant quantity $J$~\cite{jarlskog} given by
\begin{eqnarray}
J={\rm{Im}}(V_{11}V_{22}V_{12}^\ast V_{21}^\ast)=s_1^2s_2s_3c_1c_2c_3\sin{\delta}\label{jarskog}
\end{eqnarray}
is independent of phase convention, making it important when
discussing CP violation. Expanding $J$ with $\epsilon_i$ to the
second order gives
\begin{eqnarray}
J&=&J_0(1+\epsilon_1(3\cot{2\theta_1^0}+\csc{2\theta_1^0})+2\epsilon_2\cot{2\theta_2^0}+2\epsilon_3\cot{2\theta_3^0}
+\frac{1}{4}\epsilon_1^2(9\cos{2\theta_1^0}-5)\csc^2{\theta_1^0}\nonumber\\
&-&2\epsilon_2^2-2\epsilon_3^2+2\epsilon_1\epsilon_2(3\cos{2\theta_1^0}+1)\cot{2\theta_2^0}\csc{2\theta_1^0}
+4\epsilon_2\epsilon_3\cot{2\theta_2^0}\cot{2\theta_3^0}\nonumber\\
&+&2\epsilon_1\epsilon_3(3\cos{2\theta_1^0}+1)\cot{2\theta_3^0}\csc{2\theta_1^0})
+\mathcal {O}(\epsilon_i^3)\;,\label{generalj}
\end{eqnarray}
in which $J_0=(s_1^0)^2s_2^0s_3^0c_1^0c_2^0c_3^0\sin{\delta}$.

The general form Eqs.~(\ref{general}) and (\ref{generalj}) look
quite complicated since they are simple expansions in mathematics. We
can simplify the general expansion by taking
$\theta_3^0=0$ because $\theta_3^0$ is small in both quark and
lepton sectors. In this case, the result is
\begin{eqnarray}
V&=&\left( \begin{array}{ccc} c_1^0 & -s_1^0 & 0
\\s_1^0c_2^0 & c_1^0c_2^0 & s_2^0e^{i\delta}
\\s_1^0s_2^0&c_1^0s_2^0&-c_2^0e^{i\delta}
\end{array}\right)
+\epsilon_1\left( \begin{array}{ccc} -s_1^0 & -c_1^0 &0
\\ c_1^0c_2^0 & -c_2^0s_1^0 & 0\\
c_1^0s_2^0 & s_1^0s_2^0 &0
\end{array}\right)
+\epsilon_2\left(\begin{array}{ccc}
0&0&0\\
-s_1^0s_2^0 &-c_1^0s_2^0&c_2^0e^{i\delta}\\
c_2^0s_1^0 &c_1^0c_2^0&s_2^0e^{i\delta}
\end{array}\right)\nonumber\\
&+&\epsilon_3\left( \begin{array}{ccc}
0&0&-s_1^0\\
0&-s_2^0e^{i\delta}&c_1^0c_2^0\\
0&c_2^0e^{i\delta}&c_1^0s_2^0
\end{array}\right)
+\frac{1}{2}\epsilon_1^2\left(\begin{array}{ccc}
-c_1^0&s_1^0&0\\
-c_2^0s_1^0&-c_1^0c_2^0&0\\
-s_1^0s_2^0&-c_1^0s_2^0&0
\end{array}\right)
+\frac{1}{2}\epsilon_2^2\left(\begin{array}{ccc}
0&0&0\\
-c_2^0s_1^0&-c_1^0c_2^0&-s_2^0e^{i\delta}\\
-s_1^0s_2^0&-c_1^0s_2^0&c_2^0e^{i\delta}
\end{array}\right)\nonumber\\
&+&\frac{1}{2}\epsilon_3^2\left(\begin{array}{ccc}
0&s_1^0&0\\
0&-c_1^0c_2^0&-s_2^0e^{i\delta}\\
0&-c_1^0s_2^0&c_2^0e^{i\delta}
\end{array}\right)
+\epsilon_1\epsilon_2\left( \begin{array}{ccc}
0&0&0\\
-c_1^0s_2^0&s_1^0s_2^0&0\\
c_1^0c_2^0&-c_2^0s_1^0&0
\end{array}\right)
+\epsilon_2\epsilon_3\left( \begin{array}{ccc}
0&0&0\\
0&-c_2^0e^{i\delta}&-c_1^0s_2^0\\
0&-s_2^0e^{i\delta}&c_1^0c_2^0
\end{array}\right)\nonumber\\
&+&\epsilon_1\epsilon_3\left(\begin{array}{ccc}
0&0&-c_1^0\\
0&0&-c_2^0s_1^0\\
0&0&-s_1^0s_2^0
\end{array}\right)
+\mathcal {O}(\epsilon_i^3)\;,
\end{eqnarray}
and the Jarlskog parameter reduces to
\begin{eqnarray}
J=\sin{\delta}(\epsilon_3c_1^0c_2^0(s_1^0)^2s_2^0+\epsilon_2\epsilon_3c_1^0(s_1^0)^2\cos{2\theta_2^0}
-\frac{1}{8}\epsilon_1\epsilon_3\sin{2\theta_2^0}(s_1^0-3\sin{3\theta_1^0}))\;.\nonumber\\
\end{eqnarray}
These expressions for mixing matrix and Jarlskog parameter are still
complicated, making it difficult to capture physical meanings from
them. A good choice of the zeroth order matrix $V_0$ will simplify the
parametrization greatly, lead to fast convergency of the expansion,
reflect the physical insight of a parametrization, and provide
hints for underlying theory producing the mixing. Therefore, in the
following two sections, different basis matrices will be applied to
make the expansions simpler and useful for both theoretical and
experimental analysis.

\section{Parametrization of KM matrix with unit and bimaximal basis matrices}
\subsection{The triminimal expansion}
Since $V_{\rm CKM}$ is close to the unit matrix shown by
Eq.~(\ref{ckmdata}), we can naturally set
\begin{eqnarray}
\theta_1^{0Q}=\theta_2^{0Q}=\theta_3^{0Q}=0\;,\label{unitcondition}
\end{eqnarray}
and consequently have
\begin{eqnarray}
\epsilon_1^Q=\theta_1^Q=0.2273_{-0.0003}^{+0.0011},\quad
\epsilon_2^Q=\theta_2^Q=0.0383_{-0.0010}^{+0.0011},\quad
\epsilon_3^Q=\theta_3^Q=0.0154_{-0.0006}^{+0.0008}\;,
\end{eqnarray}
which show that $(\epsilon_1^Q)^2\sim\epsilon_2^Q\sim\epsilon_3^Q$. In most
cases the approximation to the second order of $\epsilon_i$ is enough. However, in order to keep
the consistency of magnitudes in the expansion, we display all terms
of $\mathcal {O}((\epsilon_1^Q)^3)$ in our parametrization, which is given by
\begin{eqnarray}
V_{\rm CKM}=\left( \begin{array}{ccc}
1-\frac{(\epsilon_1^Q)^2}{2}&\epsilon_1^Q-\frac{(\epsilon_1^Q)^3}{6}&e^{-i\delta}\epsilon_1^Q\epsilon_3^Q\\
\frac{(\epsilon_1^Q)^3}{6}-\epsilon_1^Q&1-\frac{(\epsilon_1^Q)^2}{2}&\epsilon_2^Q+e^{-i\delta}\epsilon_3^Q\\
\epsilon_1^Q\epsilon_2^Q&-\epsilon_2^Q-e^{i\delta}\epsilon_3^Q&1
\end{array}\right)
+\mathcal {O}((\epsilon_1^Q)^4)\;,\label{unitexpand}
\end{eqnarray}
where the rephasing of quark fields
\begin{eqnarray}
c\rightarrow ce^{i\pi},\quad
s\rightarrow se^{i\pi},\quad
b\rightarrow be^{i(\pi+\delta)}\label{quarkrephase}
\end{eqnarray}
is implied to make the lowest order be unit matrix.

Before moving on to lepton sector, we need to talk about the QLC relations in Eq.~(\ref{kmqlc}) here. In terms of triminimal parameters, it is natural to rewrite Eq.~(\ref{kmqlc}) as
\begin{eqnarray}
\theta_1^{0Q}+\theta_1^{0L}=\frac{\pi}{4},\quad \theta_2^{0Q}+\theta_2^{0L}=\frac{\pi}{4},\quad \theta_3^{0Q}=\theta_3^{0L}=0,\label{qlc}
\end{eqnarray}
and recognize all $\epsilon_i^Q$ and $\epsilon_i^L$ as small
deviations. Therefore, corresponding to the choice in quark
sector Eq.~(\ref{unitcondition}), we have
\begin{eqnarray}
\theta_1^{0L}=\theta_2^{0L}=\frac{\pi}{4},\quad\theta_3^{0L}=0\;,
\end{eqnarray}
in lepton sector. Similar redefinition of lepton fields
\begin{eqnarray}
\mu\rightarrow\mu e^{i\pi},\quad \nu^{\mu}\rightarrow\nu^{\mu}e^{i\pi},\quad
\nu^{\tau}\rightarrow\nu^{\tau}e^{i(\pi+\delta)}\label{leptonrephase}
\end{eqnarray}
is also implied to adjust the phases.
To the second order of $\epsilon_i^L$, lepton mixing matrix is expanded as
\begin{eqnarray}
U_{\rm PMNS}&=&\left(\begin{array}{ccc}
\frac{1}{\sqrt{2}}&\frac{1}{\sqrt{2}}&0\\
-\frac{1}{2}&\frac{1}{2}&\frac{1}{\sqrt{2}}\\
\frac{1}{2}&-\frac{1}{2}&\frac{1}{\sqrt{2}}
\end{array}\right)
+\epsilon_1^L\left(\begin{array}{ccc}
-\frac{1}{\sqrt{2}}&\frac{1}{\sqrt{2}}&0\\
-\frac{1}{2}&-\frac{1}{2}&0\\
\frac{1}{2}&\frac{1}{2}&0
\end{array}\right)
+\epsilon_2^L\left(\begin{array}{ccc}
0&0&0\\
\frac{1}{2}&-\frac{1}{2}&\frac{1}{\sqrt{2}}\\
\frac{1}{2}&-\frac{1}{2}&-\frac{1}{\sqrt{2}}
\end{array}\right)
+\epsilon_3^L\left(\begin{array}{ccc}
0&0&\frac{e^{-i\delta}}{\sqrt{2}}\\
0&-\frac{e^{i\delta}}{\sqrt{2}}&\frac{e^{-i\delta}}{2}\\
0&-\frac{e^{i\delta}}{\sqrt{2}}&-\frac{e^{-i\delta}}{2}
\end{array}\right)\nonumber\\
&+&\frac{(\epsilon_1^{L})^2}{2}\left(\begin{array}{ccc}
-\frac{1}{\sqrt{2}}&-\frac{1}{\sqrt{2}}&0\\
\frac{1}{2}&-\frac{1}{2}&0\\
-\frac{1}{2}&\frac{1}{2}&0
\end{array}\right)
+\frac{(\epsilon_2^{L})^2}{2}\left(\begin{array}{ccc}
0&0&0\\
\frac{1}{2}&-\frac{1}{2}&-\frac{1}{\sqrt{2}}\\
-\frac{1}{2}&\frac{1}{2}&-\frac{1}{\sqrt{2}}
\end{array}\right)
+\frac{(\epsilon_3^{L})^2}{2}\left(\begin{array}{ccc}
0&-\frac{1}{\sqrt{2}}&0\\
0&-\frac{1}{2}&-\frac{1}{\sqrt{2}}\\
0&\frac{1}{2}&-\frac{1}{\sqrt{2}}
\end{array}\right)\nonumber\\
&+&\frac{\epsilon_1^L\epsilon_2^L}{2}\left(\begin{array}{ccc}
0&0&0\\
1&1&0\\
1&1&0
\end{array}\right)
+\epsilon_2^L\epsilon_3^L\left(\begin{array}{ccc}
0&0&0\\
0&-\frac{e^{i\delta}}{\sqrt{2}}&-\frac{e^{-i\delta}}{2}\\
0&\frac{e^{i\delta}}{\sqrt{2}}&-\frac{e^{-i\delta}}{2}
\end{array}\right)
+\epsilon_1^L\epsilon_3^L\left(\begin{array}{ccc}
0&0&\frac{e^{-i\delta}}{\sqrt{2}}\\
0&0&-\frac{e^{-i\delta}}{2}\\
0&0&\frac{e^{-i\delta}}{2}
\end{array}\right)
+\mathcal {O}((\epsilon_i^{L})^3)\;.
\end{eqnarray}
As we expect, the zeroth order is the bimaximal matrix $U_{\rm bi}$.

\subsection{Wolfenstein-like parametrization}
We now compare the triminimal parametrization with the Wolfenstein
parametrization in quark sector. Since the original Wolfenstein
parametrization takes the same phase convention as the standard CK form~\cite{unified},
which implies different choice of parameters, especially the CP-violating phase $\delta$,
it is complicated to arrive at the Wolfenstein parametrization from the triminimal expansion of KM matrix by adjusting the phases of the fields.
Therefore, we only keep the original Wolfenstein expanding parameter $\lambda$, which satisfies
$\lambda=\sin{\theta_1^Q}\approx\epsilon_1^Q-\frac{(\epsilon_1^Q)^3}{6}$,
and introduce two new parameters with
\begin{eqnarray}
f\lambda^2=\sin{\theta_2^Q}\approx\epsilon_2^Q, \quad
h\lambda^2=\sin{\theta_3^Q}\approx\epsilon_3^Q\;.\label{fhrelation}
\end{eqnarray}
The CP-violating phase $\delta$ has clear physical meaning so we naturally keep it. Finally, by substituting them into
Eq.~(\ref{unitexpand}), we obtain a new Wolfenstein-like parametrization, given by
\begin{eqnarray}
V_{\rm CKM}=\left(\begin{array}{ccc}
1-\frac{\lambda^2}{2}&\lambda&e^{-i\delta}h\lambda^3\\
-\lambda&1-\frac{\lambda^2}{2}&(f+e^{-i\delta}h)\lambda^2\\
f\lambda^3&-(f+e^{i\delta}h)\lambda^2&1
\end{array}\right)\;,\label{newwolf}
\end{eqnarray}
which is a new simple form of quark mixing matrix given in our
recent work~\cite{qinnan}. Direct calculation of
$f\lambda^3=|V_{td}|$, $h\lambda^3=|V_{ub}|$ and
$|f+e^{-i\delta^Q}h|\lambda^2=|V_{cb}|$ with the latest
data~(\ref{ckmdata}) and~(\ref{wolfdata}) gives
\begin{eqnarray}
\lambda=0.2253\pm{0.0007},\quad h=0.303^{+0.014}_{-0.010},\quad f=0.754^{+0.022}_{-0.018},\quad \delta^Q={90.97^{\circ}}_{-4.44^{\circ}}^{+2.77^{\circ}}\;,\label{hf}
\end{eqnarray}
which are slightly different from the results in Ref.~\cite{qinnan}
where previous data are used. This new form of quark mixing matrix
preserves the hierarchical structure of the mixing. More
importantly, it is convenient for numerical analysis, especially for
constraint of the CP-violating phase. Along with unitarity
boomerangs it may be useful to study the presence of new physics.

To get unified Wolfenstein-like parametrizations for quark and
lepton mixing, we need to use the QLC relations Eq.~(\ref{kmqlc}),
of which the first two equations lead to the choice of the
parameters in the lepton sector as
\begin{eqnarray}
\theta_1^L=\pi/4-\arcsin{\lambda},
\quad\theta_2^L=\pi/4-\arcsin{f\lambda^2},
\end{eqnarray}
i.e., parameters $\lambda$ and $f$, which we introduce in quark
sector are also employed in lepton sector. The other two
parameters are one angle parameter related to $\theta_3^L$ and a
CP-violating phase $\delta^L$. Since the experimental data for the
small angle $\theta_3^L$ is not accurate enough, we can either set
$\theta_3^L=\eta\lambda$ or $\theta_3^L=\eta'\lambda^2$ depending on
the value of $|U_{e3}|$. We now discuss these two cases separately.

{\rm Case 1: $\theta_3^L=\eta\lambda$}

The lepton mixing matrix can be expanded in order of $\lambda$ as
\begin{eqnarray}
U_{\rm PMNS}&=&\left(
\begin{array}{ccc}
 \frac{1}{\sqrt{2}} & \frac{1}{\sqrt{2}} & 0 \\
 -\frac{1}{2} & \frac{1}{2} & \frac{1}{\sqrt{2}} \\
 \frac{1}{2} & -\frac{1}{2} & \frac{1}{\sqrt{2}}
\end{array}
\right) +\lambda\left(
\begin{array}{ccc}
 \frac{1}{\sqrt{2}} & -\frac{1}{\sqrt{2}} & \frac{e^{-i \delta } \eta }{\sqrt{2}} \\
 \frac{1}{2} & \frac{1}{2}-\frac{e^{i \delta } \eta }{\sqrt{2}} & \frac{1}{2} e^{-i \delta } \eta \\
 -\frac{1}{2} & -\frac{e^{i \delta } \eta }{\sqrt{2}}-\frac{1}{2} & -\frac{1}{2} e^{-i \delta } \eta
\end{array}
\right)\nonumber\\
&+&\lambda^2\left(
\begin{array}{ccc}
 -\frac{1}{2 \sqrt{2}} & -\frac{\eta ^2}{2 \sqrt{2}}-\frac{1}{2 \sqrt{2}} & -\frac{e^{-i \delta } {\eta} }{\sqrt{2}} \\
 \frac{1}{4}-\frac{f}{2} & -\frac{\eta ^2}{4}+\frac{f}{2}-\frac{1}{4} & -\frac{\eta ^2}{2 \sqrt{2}}+\frac{1}{2} e^{-i \delta } \eta -\frac{f}{\sqrt{2}} \\
 -\frac{f}{2}-\frac{1}{4} & \frac{\eta ^2}{4}+\frac{f}{2}+\frac{1}{4} & -\frac{(\eta ^2}{2 \sqrt{2}}-\frac{1}{2} e^{-i \delta } \eta +\frac{f}{\sqrt{2}}
\end{array}
\right)\nonumber\\
&+&\lambda^3\left(
\begin{array}{ccc}
 0 & \frac{\eta ^2}{2 \sqrt{2}} & -\frac{e^{-i \delta } \eta ^3}{6 \sqrt{2}}-\frac{e^{-i \delta } \eta' }{2 \sqrt{2}} \\
 \frac{f}{2} & \frac{e^{i \delta } \eta ^3}{6 \sqrt{2}}-\frac{\eta ^2}{4}+\frac{e^{i \delta } f \eta }{\sqrt{2}}+\frac{f}{2} & -\frac{1}{12} e^{-i \delta } \eta ^3-\frac{1}{4} e^{-i \delta } \eta +\frac{1}{2} e^{-i \delta } f \eta  \\
 \frac{f}{2} & \frac{e^{i \delta } \eta ^3}{6 \sqrt{2}}+\frac{\eta ^2}{4}-\frac{e^{i \delta } f \eta }{\sqrt{2}}+\frac{f}{2} & \frac{1}{12} e^{-i \delta } \eta ^3+\frac{1}{4} e^{-i \delta } \eta +\frac{1}{2} e^{-i \delta } f \eta
\end{array}
\right)
+\mathcal {O}(\lambda^4)\;.
\end{eqnarray}
With the modulus of the element $U_{e3}$ in Eq.~(\ref{pmnsdata}) and
the results for $\lambda$ and $f$ in Eq.~(\ref{hf}), the new
parameter is constrained by $0<\eta<1.923$.

{\rm Case 2: $\theta_3^L=\eta'\lambda^2$}

In this case the expansion looks simpler since $\eta'$ only starts to appear
in $\mathcal{O}(\lambda^2)$ terms:
\begin{eqnarray}
U_{\rm PMNS}&=&\left(
\begin{array}{ccc}
 \frac{1}{\sqrt{2}} & \frac{1}{\sqrt{2}} & 0 \\
 -\frac{1}{2} & \frac{1}{2} & \frac{1}{\sqrt{2}} \\
 \frac{1}{2} & -\frac{1}{2} & \frac{1}{\sqrt{2}}
\end{array}
\right) +\lambda\left(
\begin{array}{ccc}
 \frac{1}{\sqrt{2}} & -\frac{1}{\sqrt{2}} & 0 \\
 \frac{1}{2} & \frac{1}{2} & 0 \\
 -\frac{1}{2} & -\frac{1}{2} & 0
\end{array}
\right)
+\lambda^2\left(
\begin{array}{ccc}
 -\frac{1}{2 \sqrt{2}} & -\frac{1}{2 \sqrt{2}} & \frac{e^{-i \delta } \eta'}{\sqrt{2}} \\
 \frac{1}{4}-\frac{f}{2} & \frac{f}{2}-\frac{e^{i \delta } \eta'}{\sqrt{2}}-\frac{1}{4} & \frac{1}{2} e^{-i \delta } \eta'-\frac{f}{\sqrt{2}} \\
 -\frac{f}{2}-\frac{1}{4} & \frac{f}{2}-\frac{e^{i \delta } \eta'}{\sqrt{2}}+\frac{1}{4} & \frac{f}{\sqrt{2}}-\frac{1}{2} e^{-i \delta }\eta'
\end{array}
\right)\nonumber\\
&+&\lambda^3\left(
\begin{array}{ccc}
 0 & 0 & -\frac{e^{-i \delta } \eta'}{\sqrt{2}} \\
 \frac{f}{2} & \frac{f}{2} & \frac{1}{2} e^{-i \delta } \eta'\\
 \frac{f}{2} & \frac{f}{2} & -\frac{1}{2} e^{-i \delta } \eta'
\end{array}
\right)
+\mathcal {O}(\lambda^4)\;.
\end{eqnarray}
However, we have a larger range in this case with $0<\eta'<7.912$.

To the lowest order the Jarlskog parameter in both quark and lepton
sectors is given by
\begin{eqnarray}
&&J^Q={\rm{Im}}(V_{us}V_{cb}V_{ub}^\ast V_{cs}^\ast)=fh\lambda^6\sin{\delta^Q},\nonumber\\
&&J^L={\rm{Im}}(U_{e2}U_{\mu3}U_{e3}^\ast U_{\mu2}^\ast)=\frac{1}{4\sqrt{2}}\eta\lambda\sin{\delta^L},\nonumber\\
&&J^L={\rm{Im}}(U_{e2}U_{\mu3}U_{e3}^\ast U_{\mu2}^\ast)=\frac{1}{4\sqrt{2}}\eta'\lambda^2\sin{\delta^L}\;.
\end{eqnarray}
The last two equations correspond to the two cases respectively.

\section{Parametrization of the KM matrix in tri-bimaximal pattern}

\subsection{The triminimal expansion}
Present data of the PMNS matrix indicate that the mixing in the
lepton sector is closer to the tri-bimaximal mixing. One would have
a much faster convergent expansion if it starts
with a tri-bimaximal mixing form as the basis, which implies again the rephasing of lepton fields in Eq.~(\ref{leptonrephase}) and the choice of $\theta_i^{0L}$ as
\begin{eqnarray}
\theta_1^{0L}=\arcsin{\frac{1}{\sqrt{3}}},\quad\theta_2^{0L}=\frac{\pi}{4},\quad\theta_3^{0L}=0\;.
\end{eqnarray}
Under such conditions, the lepton mixing matrix expanded to the second order of $\epsilon_i^L$ is given by
\begin{eqnarray}
U_{\rm PMNS}&=&\left(\begin{array}{ccc}
\sqrt{\frac{2}{3}}&\frac{1}{\sqrt{3}}&0\\
-\frac{1}{\sqrt{6}}&\frac{1}{\sqrt{3}}&\frac{1}{\sqrt{2}}\\
\frac{1}{\sqrt{6}}&-\frac{1}{\sqrt{3}}&\frac{1}{\sqrt{2}}
\end{array}\right)
+\epsilon_1^L\left(\begin{array}{ccc}
-\frac{1}{\sqrt{3}}&\sqrt{\frac{2}{3}}&0\\
-\frac{1}{\sqrt{3}}&-\frac{1}{\sqrt{6}}&0\\
\frac{1}{\sqrt{3}}&\frac{1}{\sqrt{6}}&0
\end{array}\right)
+\epsilon_2^L\left(\begin{array}{ccc}
0&0&0\\
\frac{1}{\sqrt{6}}&-\frac{1}{\sqrt{3}}&\frac{1}{\sqrt{2}}\\
\frac{1}{\sqrt{6}}&-\frac{1}{\sqrt{3}}&-\frac{1}{\sqrt{2}}
\end{array}\right)\nonumber\\
&+&\epsilon_3^L\left(\begin{array}{ccc}
0&0&\frac{e^{-i\delta}}{\sqrt{3}}\\
0&-\frac{e^{i\delta}}{\sqrt{2}}&\frac{e^{-i\delta}}{\sqrt{3}}\\
0&-\frac{e^{i\delta}}{\sqrt{2}}&-\frac{e^{-i\delta}}{\sqrt{3}}
\end{array}\right)
+\frac{(\epsilon_1^L)^2}{2}\left(\begin{array}{ccc}
-\sqrt{\frac{2}{3}}&-\frac{1}{\sqrt{3}}&0\\
\frac{1}{\sqrt{6}}-\frac{1}{\sqrt{3}}&0\\
-\frac{1}{\sqrt{6}}&\frac{1}{\sqrt{3}}&0
\end{array}\right)
+\frac{(\epsilon_2^L)^2}{2}\left(\begin{array}{ccc}
0&0&0\\
\frac{1}{\sqrt{6}}&-\frac{1}{\sqrt{3}}&-\frac{1}{\sqrt{2}}\\
-\frac{1}{\sqrt{6}}&\frac{1}{\sqrt{3}}&-\frac{1}{\sqrt{2}}
\end{array}\right)\nonumber\\
&+&\frac{(\epsilon_3^L)^2}{2}\left(\begin{array}{ccc}
0&-\frac{1}{\sqrt{3}}&0\\
0&-\frac{1}{\sqrt{3}}&-\frac{1}{\sqrt{2}}\\
0&\frac{1}{\sqrt{3}}&-\frac{1}{\sqrt{2}}
\end{array}\right)
+\epsilon_1^L\epsilon_2^L\left(\begin{array}{ccc}
0&0&0\\
\frac{1}{\sqrt{3}}&\frac{1}{\sqrt{6}}&0\\
\frac{1}{\sqrt{3}}&\frac{1}{\sqrt{6}}&0
\end{array}\right)
+\epsilon_2^L\epsilon_3^L\left(\begin{array}{ccc}
0&0&0\\
0&-\frac{e^{i\delta}}{\sqrt{2}}&-\frac{e^{-i\delta}}{\sqrt{3}}\\
0&\frac{e^{i\delta}}{\sqrt{2}}&-\frac{e^{-i\delta}}{\sqrt{3}}
\end{array}\right)\nonumber\\
&+&\epsilon_1^L\epsilon_3^L\left(\begin{array}{ccc}
0&0&\sqrt{\frac{2}{3}}e^{-i\delta}\\
0&0&-\frac{e^{-i\delta}}{\sqrt{6}}\\
0&0&\frac{e^{-i\delta}}{\sqrt{6}}
\end{array}\right)
+\mathcal {O}((\epsilon_i^L)^3)\;.
\end{eqnarray}

With Eq.~(\ref{qlc}) as our guide, we get the zeroth order of mixing
angles in the quark sector as
\begin{eqnarray}
\theta_1^{0Q}=\arcsin{\frac{\sqrt{2}-1}{\sqrt{6}}},~\theta_2^{0Q}=0,~\theta_3^{0Q}=0\;.
\end{eqnarray}
In this case the deviations are
\begin{eqnarray}
\epsilon_1^Q=0.0574_{-0.0003}^{+0.0010},\quad
\epsilon_2^Q=0.0383_{-0.0010}^{+0.0011},\quad
\epsilon_3^Q=0.0154_{-0.0006}^{+0.0008}\;,
\end{eqnarray}
according to Eq.~(\ref{kmdata}).
Thus $\epsilon_i^Q\sim\mathcal {O}(10^{-2})$, and we have a faster
convergent expansion for quark mixing as
\begin{eqnarray}
V_{\rm CKM}&=&\left(\begin{array}{ccc}
\frac{\sqrt{2}+1}{\sqrt{6}}&\frac{\sqrt{2}-1}{\sqrt{6}}&0\\
-\frac{\sqrt{2}-1}{\sqrt{6}}&\frac{\sqrt{2}+1}{\sqrt{6}}&0\\
0&0&1
\end{array}\right)
+\epsilon_1^Q\left(\begin{array}{ccc}
-\frac{\sqrt{2}-1}{\sqrt{6}}&\frac{\sqrt{2}+1}{\sqrt{6}}&0\\
-\frac{\sqrt{2}+1}{\sqrt{6}}&-\frac{\sqrt{2}-1}{\sqrt{6}}&0\\
0&0&0
\end{array}\right)
+\epsilon_2^Q\left(\begin{array}{ccc}
0&0&0\\
0&0&1\\
\frac{\sqrt{2}-1}{\sqrt{6}}&-\frac{\sqrt{2}+1}{\sqrt{6}}&0
\end{array}\right)\nonumber\\
&+&\epsilon_3^Q\left(\begin{array}{ccc}
0&0&\frac{\sqrt{2}-1}{\sqrt{6}}e^{-i\delta}\\
0&0&\frac{\sqrt{2}+1}{\sqrt{6}}e^{-i\delta}\\
0&-e^{i\delta}&0
\end{array}\right)
+\frac{(\epsilon_1^Q)^2}{2}\left(\begin{array}{ccc}
-\frac{\sqrt{2}+1}{\sqrt{6}}&-\frac{\sqrt{2}-1}{\sqrt{6}}&0\\
\frac{\sqrt{2}-1}{\sqrt{6}}&-\frac{\sqrt{2}+1}{\sqrt{6}}&0\\
0&0&0
\end{array}\right)
+\frac{(\epsilon_2^Q)^2}{2}\left(\begin{array}{ccc}
0&0&0\\
\frac{\sqrt{2}-1}{\sqrt{6}}&-\frac{\sqrt{2}+1}{\sqrt{6}}&0\\
0&0&-1
\end{array}\right)\nonumber\\
&+&\frac{(\epsilon_3^Q)^2}{2}\left(\begin{array}{ccc}
0&-\frac{\sqrt{2}-1}{\sqrt{6}}&0\\
0&-\frac{\sqrt{2}+1}{\sqrt{6}}&0\\
0&0&-1
\end{array}\right)
+\epsilon_1^Q\epsilon_2^Q\left(\begin{array}{ccc}
0&0&0\\
0&0&0\\
\frac{\sqrt{2}+1}{\sqrt{6}}&\frac{\sqrt{2}-1}{\sqrt{6}}&0
\end{array}\right)
+\epsilon_2^Q\epsilon_3^Q\left(\begin{array}{ccc}
0&0&0\\
0&-e^{i\delta}&0\\
0&0&-\frac{\sqrt{2}+1}{\sqrt{6}}e^{-i\delta}
\end{array}\right)\nonumber\\
&+&\epsilon_1^Q\epsilon_3^Q\left(\begin{array}{ccc}
0&0&\frac{\sqrt{2}+1}{\sqrt{6}}e^{-i\delta}\\
0&0&-\frac{\sqrt{2}-1}{\sqrt{6}}e^{-i\delta}\\
0&0&0
\end{array}\right)
+\mathcal {O}((\epsilon_i^Q)^3)\;.
\end{eqnarray}
Consistent with the results in Ref.~\cite{lisw}, the QLC relations
relate the tri-bimaximal matrix in the lepton sector with $V'_0$ in
the quark sector, noting that they are both more close to
experimental data compared with the bimaximal matrix and unit
matrix.

\subsection{Wolfenstein-like parametrization}
Similarly with Sec.~III, we now discuss the corresponding Wolfenstein-like parametrizations in both quark sector and lepton sector in a unified way with the help of QLC relations.
For quark mixing, we can employ $\rho=\epsilon_1^Q$ as expanding parameter and introduce two coefficients with $s\rho=\epsilon_2^Q$, $t\rho=\epsilon_3^Q$, i.e.,
\begin{eqnarray}
\theta_1^Q=\arcsin{\frac{\sqrt{2}-1}{\sqrt{6}}}+\rho,\quad\theta_2^Q=s\rho,\quad\theta_3^Q=t\rho\;.
\end{eqnarray}
Since the expansion around basis $V'_0$ converges faster than the
case in Sec.~III, it is accurate enough to calculate to $\mathcal {O}(\rho^2)$,
given by
\begin{eqnarray}
V_{\rm CKM}&=&\left(\begin{array}{ccc}
\frac{\sqrt{2}+1}{\sqrt{6}}&\frac{\sqrt{2}-1}{\sqrt{6}}&0\\
-\frac{\sqrt{2}-1}{\sqrt{6}}&\frac{\sqrt{2}+1}{\sqrt{6}}&0\\
0&0&1
\end{array}\right)
+\rho\left(
\begin{array}{ccc}
 -\frac{\sqrt{2}-1}{\sqrt{6}} & \frac{\sqrt{2}+1}{\sqrt{6}} & \frac{(\sqrt{2}-1)e^{-i\delta}t}{\sqrt{6}} \\
 -\frac{\sqrt{2}+1}{\sqrt{6}} & -\frac{\sqrt{2}-1}{\sqrt{6}} & s+\frac{(\sqrt{2}+1)e^{-i\delta}t}{\sqrt{6}} \\
 \frac{(\sqrt{2}-1)s}{\sqrt{6}} & -\frac{(\sqrt{2}+1)s}{\sqrt{6}}-e^{i\delta}t & 0
\end{array}
\right)\nonumber\\
&+&\rho^2\left(
\begin{array}{ccc}
 -\frac{\sqrt{2}+1}{2\sqrt{6}} & -\frac{(\sqrt{2}-1)(t^2+1)}{2\sqrt{6}} & \frac{(\sqrt{2}+1)e^{-i\delta}t}{\sqrt{6}} \\
 \frac{(\sqrt{2}-1)(s^2+1)}{2\sqrt{6}} & -\frac{(\sqrt{2}+1)(s^2+t^2+1)}{2\sqrt{6}}-e^{i\delta}ts & -\frac{(\sqrt{2}-1)e^{-i\delta}t}{\sqrt{6}} \\
 \frac{(\sqrt{2}+1)s}{\sqrt{6}} & \frac{(\sqrt{2}-1)s}{\sqrt{6}} & -\frac{(\sqrt{2}+1)e^{-i\delta}ts}{\sqrt{6}}-\frac{s^2}{2}-\frac{t^2}{2}
\end{array}
\right)\nonumber\\
&+&\mathcal {O}(\rho^3)\;.
\end{eqnarray}
By using the data for elements $V_{ud}$, $V_{ub}$, $V_{td}$ and $V_{cb}$ in Eq.~(\ref{ckmdata}), we get the ranges for the
parameters with
\begin{eqnarray}
\rho=0.0574^{+0.0006}_{-0.0007},\quad s=0.667_{-0.016}^{+0.020},\quad t=0.268_{-0.002}^{+0.013},\quad \delta^Q={90.30^{\circ}}_{-4.57^{\circ}}^{+2.85^{\circ}}\;.
\end{eqnarray}
As we can see here, the expanding parameter is $\rho\sim\mathcal
{O}(10^{-2})$, making the expansion converge fast as we mentioned
before, and indicating that $V'_0$ is indeed a good choice of basis matrix.

For the corresponding Wolfenstein-like matrix in lepton sector, by using the QLC relations in Eq.~(\ref{kmqlc}),
we have
\begin{eqnarray}
\theta_1^L=\frac{\pi}{4}-(\arcsin{\frac{\sqrt{2}-1}{\sqrt{6}}}+\rho),\quad\theta_2^L=\frac{\pi}{4}-s\rho\;.
\end{eqnarray}
We still need to retain the CP-violating phase $\delta^L$ and to introduce a new parameter $\tau$ (or $\tau'$). Here similar discussions as in Sec.~III are needed depending on the
value of $|U_{e3}|$.

{\rm Case 1: $\theta_3^L=\tau\rho$}

To good accuracy, we expand the PMNS matrix to $\mathcal
{O}(\rho^{2})$ and obtain
\begin{eqnarray}
U_{\rm PMNS}&=&\left(\begin{array}{ccc}
\sqrt{\frac{2}{3}}&\frac{1}{\sqrt{3}}&0\\
-\frac{1}{\sqrt{6}}&\frac{1}{\sqrt{3}}&\frac{1}{\sqrt{2}}\\
\frac{1}{\sqrt{6}}&-\frac{1}{\sqrt{3}}&\frac{1}{\sqrt{2}}
\end{array}\right)
+\rho\left(
\begin{array}{ccc}
 \frac{1}{\sqrt{3}} & -\sqrt{\frac{2}{3}} & \frac{e^{-i \delta } \tau }{\sqrt{3}} \\
 \frac{1}{\sqrt{3}}-\frac{s}{\sqrt{6}} & \frac{s}{\sqrt{3}}-\frac{e^{i \delta } \tau }{\sqrt{2}}+\frac{1}{\sqrt{6}} & \frac{e^{-i \delta } \tau }{\sqrt{3}}-\frac{s}{\sqrt{2}} \\
 -\frac{s}{\sqrt{6}}-\frac{1}{\sqrt{3}} & \frac{s}{\sqrt{3}}-\frac{e^{i \delta } \tau }{\sqrt{2}}-\frac{1}{\sqrt{6}} & \frac{s}{\sqrt{2}}-\frac{e^{-i \delta } \tau }{\sqrt{3}}
\end{array}
\right)\nonumber\\
&+&\rho^2\left(
\begin{array}{ccc}
 -\frac{1}{\sqrt{6}} & -\frac{\tau ^2}{2 \sqrt{3}}-\frac{1}{2 \sqrt{3}} & -\sqrt{\frac{2}{3}} e^{-i \delta } \tau  \\
 \frac{s^2}{2 \sqrt{6}}+\frac{s}{\sqrt{3}}+\frac{1}{2 \sqrt{6}} & -\frac{s^2}{2 \sqrt{3}}+\frac{e^{i \delta } \tau  s}{\sqrt{2}}+\frac{s}{\sqrt{6}}-\frac{\tau ^2}{2 \sqrt{3}}-\frac{1}{2 \sqrt{3}} & -\frac{s^2}{2 \sqrt{2}}+\frac{e^{-i \delta } \tau  s}{\sqrt{3}}-\frac{\tau ^2}{2 \sqrt{2}}+\frac{e^{-i \delta } \tau }{\sqrt{6}} \\
 -\frac{s^2}{2 \sqrt{6}}+\frac{s}{\sqrt{3}}-\frac{1}{2 \sqrt{6}} & \frac{s^2}{2 \sqrt{3}}-\frac{e^{i \delta } \tau  s}{\sqrt{2}}+\frac{s}{\sqrt{6}}+\frac{\tau ^2}{2 \sqrt{3}}+\frac{1}{2 \sqrt{3}} & -\frac{s^2}{2 \sqrt{2}}+\frac{e^{-i \delta } \tau  s}{\sqrt{3}}-\frac{\tau ^2}{2 \sqrt{2}}-\frac{e^{-i \delta } \tau }{\sqrt{6}}
\end{array}
\right)\nonumber\\
&+&\mathcal {O}(\rho^3)\;.
\end{eqnarray}
The upper bound of $|U_{e3}|$ gives $0<\tau<7.493$.

{\rm Case 2: $\theta_3^L=\tau'\rho^2$}

The expansion to $\mathcal {O}(\rho^2)$ is given by
\begin{eqnarray}
U_{\rm PMNS}&=&\left(\begin{array}{ccc}
\sqrt{\frac{2}{3}}&\frac{1}{\sqrt{3}}&0\\
-\frac{1}{\sqrt{6}}&\frac{1}{\sqrt{3}}&\frac{1}{\sqrt{2}}\\
\frac{1}{\sqrt{6}}&-\frac{1}{\sqrt{3}}&\frac{1}{\sqrt{2}}
\end{array}\right)
+\rho\left(
\begin{array}{ccc}
 \frac{1}{\sqrt{3}} & -\sqrt{\frac{2}{3}} & 0 \\
 \frac{1}{\sqrt{3}}-\frac{s}{\sqrt{6}} & \frac{s}{\sqrt{3}}+\frac{1}{\sqrt{6}} & -\frac{s}{\sqrt{2}} \\
 -\frac{s}{\sqrt{6}}-\frac{1}{\sqrt{3}} & \frac{s}{\sqrt{3}}-\frac{1}{\sqrt{6}} & \frac{s}{\sqrt{2}}
\end{array}
\right)\nonumber\\
&+&\rho^2\left(
\begin{array}{ccc}
 -\frac{1}{\sqrt{6}} & -\frac{1}{2 \sqrt{3}} & \frac{e^{-i \delta } \tau'}{\sqrt{3}} \\
 \frac{s^2}{2 \sqrt{6}}+\frac{s}{\sqrt{3}}+\frac{1}{2 \sqrt{6}} & -\frac{s^2}{2 \sqrt{3}}+\frac{s}{\sqrt{6}}-\frac{e^{i \delta } \tau' }{\sqrt{2}}-\frac{1}{2\sqrt{3}} & \frac{e^{-i \delta } \tau' }{\sqrt{3}}-\frac{s^2}{2 \sqrt{2}} \\
 -\frac{s^2}{2 \sqrt{6}}+\frac{s}{\sqrt{3}}-\frac{1}{2 \sqrt{6}} & \frac{s^2}{2 \sqrt{3}}+\frac{s}{\sqrt{6}}-\frac{e^{i \delta } \tau' }{\sqrt{2}}+\frac{1}{2\sqrt{3}} & -\frac{s^2}{2 \sqrt{2}}-\frac{e^{-i \delta } \tau'}{\sqrt{3}}
\end{array}
\right)\nonumber\\
&+&\mathcal {O}(\rho^3)\;.
\end{eqnarray}
A larger range for the parameter is obtained as $0<\tau'<125.987$.
If we would like to control the parameters to be $\mathcal
{O}(10^{-1})$, the former case is a better choice. However, further neutrino oscillation experiments are needed to determine
$|U_{e3}|$ and the parameters we adopt here.

The Jarlskog parameters in tri-bimaximal pattern are given by
\begin{eqnarray}
&&J^Q={\rm{Im}}(V_{us}V_{cb}V_{ub}^\ast V_{cs}^\ast)=\frac{1}{6}\sqrt{\frac{1}{6}(3-2\sqrt{2})}st\rho^2\sin{\delta^Q},\nonumber\\
&&J^L={\rm{Im}}(U_{e2}U_{\mu3}U_{e3}^\ast U_{\mu2}^\ast)=\frac{\tau\rho\sin{\delta^L}}{3\sqrt{6}},\nonumber\\
&&J^L={\rm{Im}}(U_{e2}U_{\mu3}U_{e3}^\ast U_{\mu2}^\ast)=\frac{\tau'\rho^2\sin{\delta^L}}{3\sqrt{6}}\;.
\end{eqnarray}

\section{Discussions and conclusions}
In this paper, by using triminimal and Wolfenstein-like expansion
techniques, we study in more detail the Kobayashi-Maskawa matrix in
both quark and lepton sectors. Our motivation is based on the
consideration of the convenience the KM form exhibits when
discussing some problems such as unitarity boomerang and maximal CP
violation.

In the previous two sections, we choose the unit matrix and $V'_0$
as basis matrices for quark mixing, while bimaximal and
tri-bimaximal matrices for lepton mixing. Naturally, a question
arises here: which pattern is better, bimaximal, or tri-bimaximal? On
one hand, the bimaximal matrix is related with the unit matrix through
QLC relations. The corresponding triminimal and Wolfenstein-like
expansions are both comparatively simple and symmetric. On the other
hand, compared with bimaximal matrix, the tri-bimaximal matrix is closer
to the experimental data; thus, the expansion based on it converges
much faster. This can be reflected from the smallness of the
triminimal angles and the Wolfenstein-like expanding parameter.
Therefore, if we are interested only in the leading order
contribution, then the tri-bimaximal matrix should be chosen. Theoretically,
there have been some attempts at understanding both bimaximal
and tri-bimaximal matrices with the introduction of new symmetries
to the standard model fermions. However, we still need a more fundamental
theory to decide which pattern should be used, to reveal the origin of the
mixing matrices.

The Wolfenstein-like parametrizations presented in this paper are simpler than the
results in Ref.~\cite{unified} and can be transformed to triminimal
expansions with relations among parameters. Here by using two sets of
the Wolfenstein-like parameters, i.e., $\lambda$, $f$, $h$, $\eta$(or
$\eta'$), $\delta^Q$, and $\delta^L$ in Sec.~III and $\rho$, $s$,
$t$, $\tau$(or $\tau'$), $\delta^Q$, and $\delta^L$ in Sec.~IV, we
unify the parametrization of the KM matrix in quark and lepton sectors.
The parameters in the quark sector can be well determined with
current experimental data. However, the experimental results in
lepton sector, especially for $|U_{e3}|$ and the CP-violating phase
$\delta^L$, are far from enough.

The unified description of fermion mixing we get here results from
the QLC relations in the KM form, and these relations are only
approximately valid; thus, one may doubt the validity of the results.
Actually, we can discuss this in a reversed way. By regarding these
parameters for quarks and leptons as independent from each other, we
can determine the parameters with data from quark experiments and
lepton oscillations separately, and check the QLC relations. Even if
the QLC relations violate, these parametrizations still stand
separately in the quark sector and lepton sector. Thus, our
study is helpful in understanding the mixing phenomenologically and
may provide useful tools in searching for a profound theory on the fermion
masses and mixing.

\begin{acknowledgments}
This work is partially supported by National Natural Science
Foundation of China (Grants No.~11021092, No.~10975003, No.~11035003, ) and by the
Key Grant Project of Chinese Ministry of Education (Grant No.~305001).
\end{acknowledgments}

\end{document}